\documentclass[12pt]{article}

\usepackage{scicite}

\usepackage{times}

\usepackage{epsfig}
\usepackage{color}
\usepackage{graphics}
\usepackage{tabularx}
\usepackage{longtable}
\usepackage{hyperref}           
\usepackage[normalem]{ulem}

\def\tol#1#2#3{\hbox{\rule{0pt}{15pt}${#1}^{+{#2}}_{-{#3}}$}}

\newcommand{\Ms}{\mathrm{M}_\odot}

\newenvironment{sciabstract}{%
\begin{quote} \bf}
{\end{quote}}

\newcounter{lastnote}

\title{Local Group galaxies emerge from the dark}

\author {\parbox{\textwidth}{Till~Sawala$^{1\ast}$,
     Carlos~S.~Frenk$^1$,  Azadeh
  Fattahi$^2$, Julio~F.~Navarro$^2$, Richard~G.~Bower$^1$,
  Robert~A.~Crain$^3$, Claudio Dalla Vecchia$^{4,5}$, Michelle
  Furlong$^1$, John.~C.~Helly$^1$, Adrian Jenkins$^1$,
  Kyle~A.~Oman$^2$, Matthieu Schaller$^1$, Joop
  Schaye$^3$, Tom Theuns$^{1,6}$, James Trayford$^1$ and \mbox{Simon}~D.~M.~White$^7$}\vspace{0.4cm}\\
\parbox{\textwidth}{ \normalsize{
$^{1}$Institute for Computational Cosmology, Department of Physics,
University of Durham, South Road, Durham DH13LE, UK, 
$^{2}$Department of Physics and Astronomy, University of Victoria,
3800 Finnerty Road, Victoria, British Columbia V8P 5C2, Canada, 
$^{3}$Leiden Observatory, Leiden University, Postbus  9513, 2300 RA
Leiden, The Netherlands, 
$^{4}$Instituto de Astrof\'{i}sica de Canarias, C/ V\'{i}a L\'{a}ctea
s/n, 38205 La Laguna, Tenerife, Spain, 
$^{5}$Astrophysics Research Institute, Liverpool John Moores
University, 146 Brownlow Hill, Liverpool L3 5RF, UK,  
$^{6}$Department of  Physics, University of Antwerp, Campus
Groenenborger, Groenenborgerlaan 171, B-2020 Antwerp, Belgium 
$^{7}$Max Planck Institute for Astrophysics, Karl-Schwarzschild-Str. 1, 85741 Garching, Germany
\\
\\
\normalsize{$^\ast$To whom correspondence should be addressed; E-mail:  till.sawala@durham.ac.uk}}
 }}

\date{}

\begin{document} 



\maketitle 

\begin{sciabstract}
  The ``Lambda Cold Dark Matter'' ($\Lambda$CDM) model of cosmic
  structure formation is eminently falsifiable: once its parameters
  are fixed on large scales, it becomes testable in the nearby
  Universe. Observations within our Local Group of galaxies, including
  the satellite populations of the Milky Way and Andromeda, appear to
  contradict $\Lambda$CDM predictions: there are far fewer satellite
  galaxies than dark matter halos (the ``missing satellites'' problem
  \cite{Klypin-1999,Moore-1999}), galaxies seem to avoid the largest
  substructures (the ``too big to fail'' problem
  \cite{Boylan-Kolchin-2011,Ferrero-2012}), and the brightest
  satellites appear to orbit their host galaxies on a thin plane (the
  ``planes of satellites'' problem \cite{Pawlowski-2013}). We present
  results from the first hydrodynamic simulations of the Local Group
  that match the observed abundance of galaxies. We find that when
  baryonic and dark matter are followed simultaneously in the context
  of a realistic galaxy formation model, all three ``problems'' are
  resolved within the $\Lambda$CDM paradigm.
\end{sciabstract}

The ability of the cold dark matter model to predict observables on
different scales and at different epochs lies at the root of its
remarkable success. The anisotropy of the microwave background
radiation and the large scale distribution of galaxies were predicted
after the model was formulated, and have since been spectacularly
validated by observations. However, observations on scales currently
testable only within the Local Group (LG) have yielded results that no
simulation to date has been able to reproduce. This has renewed
interest in alternatives to $\Lambda$CDM, such as warm
\cite{Lovell-2012} or self-interacting \cite{Rocha-2013} dark matter.

The problems reported as fatal for $\Lambda$CDM arise when
observations are confronted with predictions from dark matter only
(DMO) simulations that treat the cosmic matter content as a single
collisionless fluid \cite{Boylan-Kolchin-2011, Garrison-Kimmel-2014},
under the assumption that baryonic processes are unimportant. It has,
of course, long been recognized that the distribution of light is not
a precise tracer of dark matter, but simple models for populating dark
matter structures with galaxies do not capture the complexity of
galaxy formation physics, including the effect of baryons on the dark
matter. On the other hand, full hydrodynamic simulations have
previously either focussed on individual galaxies ignoring the LG
environment, or have not yet been able to reproduce the LG dwarf
galaxy population \cite{Benitez-Llambay-2014, Nuza-2014}.

Improvements in numerical techniques and in the modeling of
astrophysical processes have recently lead to hydrodynamical
simulations that are able to reproduce the observed galaxy population
over large scales in cosmologically representative volumes, notably
the {\sc Illustris} \cite{Vogelsberger-2014} and {\sc Eagle}
\cite{Schaye-2014} projects. Here, we test the small scale nature of
the $\Lambda$CDM model with a new suite of cosmological hydrodynamical
simulations starting from initial conditions tailored to match the LG
environment, using the {\sc Eagle} hydrodynamics code. Applying state-
of-the-art hydrodynamic simulation techniques to a realistic local
environment allows us for the first time directly to confront
$\Lambda$CDM predictions with observations in the critical,
subgalactic regime.

In particular, we focus on pairs of halos that match the separation,
approach velocity, and relative tangential velocity of the Milky Way
(MW) and Andromeda (M31). From a large cosmological simulation, we
have selected twelve pairs of halos with combined virial masses of
$\sim2.3\pm0.6\times 10^{12} \Ms$, in agreement with the most recent
dynamical constraints \cite{Gonzalez-2013, Penarrubia-2014}. Each
region was resimulated at three different resolution levels (labelled
``L3'', ``L2'' and ``L1''), both as pure dark matter, and with the
full hydrodynamic model. At our lowest, ``L3'', resolution, the
particle mass is comparable to the intermediate resolution {\small
  EAGLE} runs, while L2 and L1 improve on this by factors of 12 and
144, respectively. At the highest resolution, each of the main
galaxies is simulated with more than $20$ million particles,
comparable to the highest resolution simulations of individual
galaxies published to date. Further details can be found in the
Supplementary Information. \\

\begin{figure*}
\hspace{-.8cm}\resizebox{17.7cm}{!}{\includegraphics{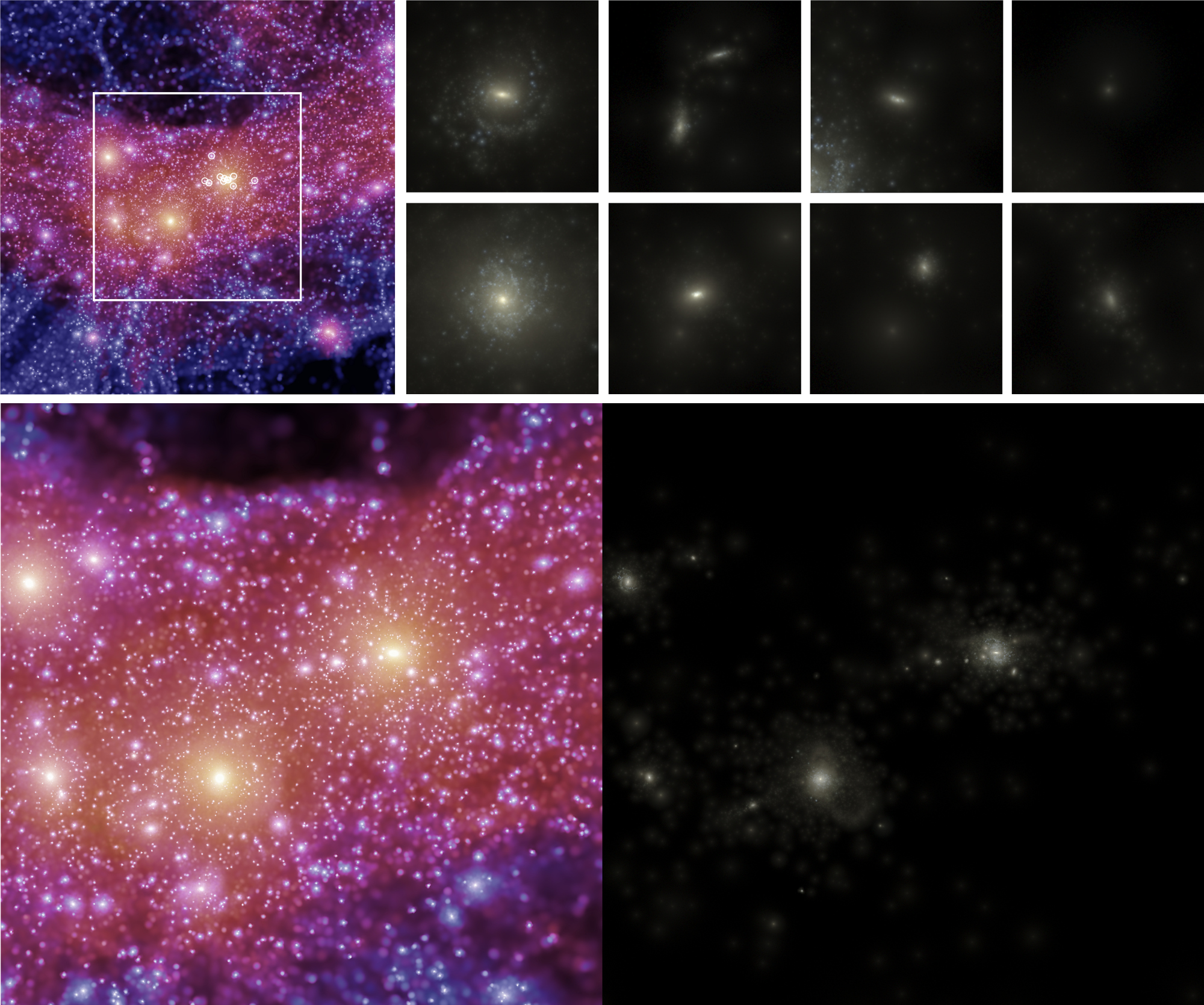}}
\caption{Distribution of dark matter and stars in a Local Group
  simulation. The top left panel shows the projected dark matter
  density distribution in one of our resimulations at resolution level
  L2 in a cube of 4~Mpc on a side. Circles indicate the locations of
  the eleven brightest satellites of one of the main galaxies, whose
  spatial distribution is as anisotropic as that of the eleven
  brightest Milky Way satellites, and which are aligned with the
  filament that contains most of the halos and galaxies in the
  region. The main panels contrast the vast number of dark matter
  substructures (left) with the stellar light distribution (right) in
  a cube of 2~Mpc, as indicated by the white square in the top left
  panel. The small panels in the top row are of side length 125 kpc
  and reveal in more detail the stellar component of some of the
  different types of galaxies formed in this simulation; central
  galaxies (first and second column) and satellite galaxies (third and
  fourth column).} \label{FigLGImage}
\end{figure*}

Dark Matter substructures are abundant in our Local Group simulations,
but due to the complexity of galaxy formation, starlight paints a very
different picture. Fig.~\ref{FigLGImage} shows the dark matter and
starlight in one of our simulation volumes at redshift $z=0$: galaxies
appear as biased tracers of the dark matter, forming almost
exclusively in the most massive halos. The top left panel shows the
dark matter distribution on larger scales, revealing a cosmic filament
that envelopes the two principal halos and most of the galaxies in the
region. Also highlighted are the positions of the halos that host the
eleven brightest satellites of one of the main halos. Analogous to the
Milky Way satellite system, the alignment is indicative of a thin
plane seen in projection, that is also aligned with the orientation of
the filament.

The small insets in Fig.~\ref{FigLGImage} show the stellar structure
of some of the many galaxies formed in this simulation. The images use
multi-band colors rendered using a spectrophotometric model. A
variety of disk and spheroid morphologies, luminosities, colors, and
sizes are clearly visible, reminiscent of the diversity of observed LG
galaxies. \\

\begin{figure}
\vspace{-0.5cm} 

\hspace{-0.8cm} \resizebox{16cm}{!}{\includegraphics{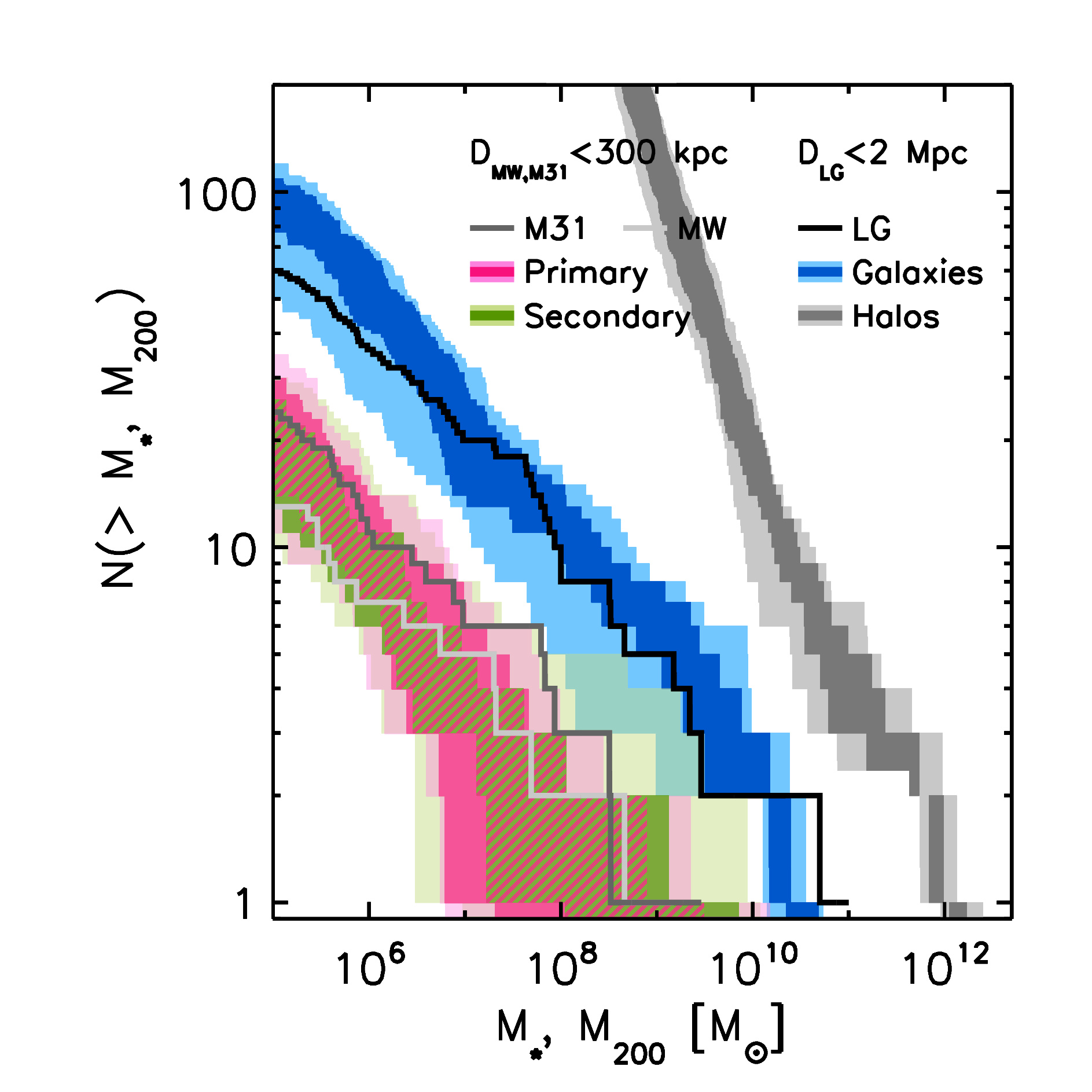}}%
\caption{Stellar and halo mass functions in the LG simulations. Red
  and green bands show the stellar mass functions of satellites within
  300 kpc of each of the two main Local Group galaxies, while blue
  lines show the number of galaxies within $2$~Mpc of the Local Group
  barycenter. The dark color-shaded areas bound the 16th and 84th
  percentiles; light shaded areas indicate the full range among our
  twelve Local Group realizations at resolution level L2. For
  comparison, the grey area corresponds to the mass function of all
  dark matter halos within $2$~Mpc. Solid lines show the measured
  stellar mass function of the satellites of the Milky Way (light
  grey) and M31 (dark grey), and of every known galaxy within $2$~Mpc
  from the Local Group barycenter (black,
  \cite{McConnachie-2012}). Note that while the M31 satellite count is
  likely to be complete to $10^5\Ms$, the count of satellites of the
  MW and the total count within $2$~Mpc should be considered as lower
  limits to the true numbers due to the limited sky coverage and the
  low surface brightness of dwarf galaxies.}  \label{FigSMF}
\end{figure}

Fig.~\ref{FigSMF} shows the galaxy stellar mass functions in the
simulations, both within $300$ kpc from each of the two main galaxies
(labelled ``primary'' and ``secondary'' in order of halo mass), as
well as within $2$ Mpc from the LG barycenter. On average, primary and
secondary galaxies have $\tol{20}{10}{6}$ and $\tol{18}{8}{5}$
satellites more massive than $M_*=10^5\Ms$ inside $300$ kpc,
respectively, in good agreement with the observed number of MW and M31
satellites\footnote{Observed stellar mass functions compiled from data
  by McConnachie \cite{McConnachie-2012}.}. Within $2$ Mpc of the LG
barycenter, there are $\sim60$ galaxies presently known with
$M_*>10^5\Ms$ known: on average, our simulations produce
$\tol{90}{20}{15}$. The modest number of luminous galaxies is in stark
contrast to the very large number of dark matter halos found within
the same volume, indicated by the grey shaded area in
Fig.~\ref{FigSMF}. While feedback from supernovae and stellar winds
regulate star formation in those halos where a dwarf galaxy has
formed, reionization has left most of the low mass halos completely
dark \cite{Sawala-2014a}.

That the simulations reproduce the stellar mass function of galaxies
and satellites in the LG over all resolved mass scales is remarkable,
given that these simulations use the very same parameters that match
the shape and normalization of the galaxy stellar mass function in
large cosmological volumes. Not only are our simulations free of the
``missing satellites'' problem, but the observed stellar mass
functions of the LG environment and of the MW and M31 satellites
are entirely consistent with $\Lambda$CDM. \\

\begin{figure}
\vspace{-1cm}
\hspace{-.8cm}
\resizebox{16cm}{!}{\includegraphics{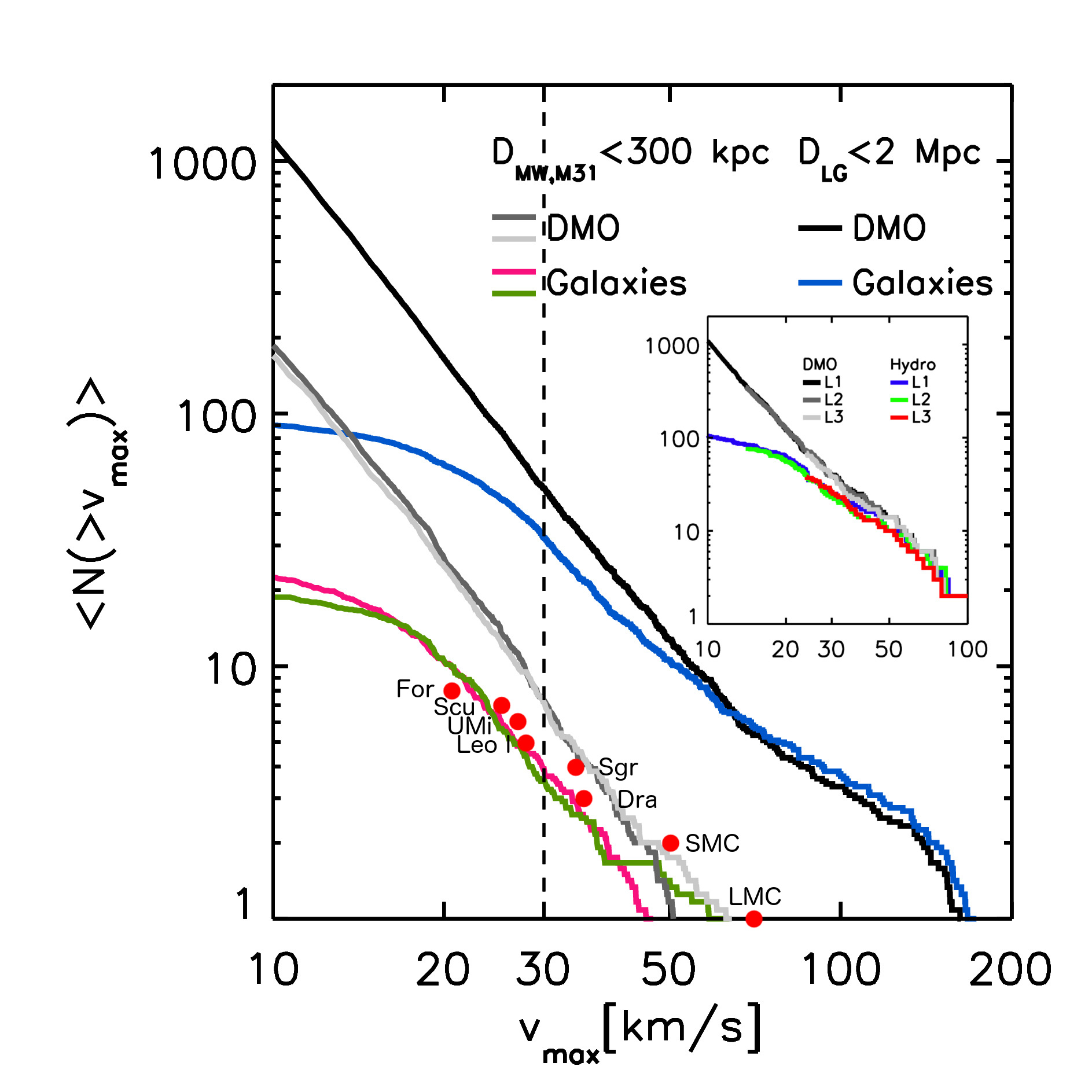}}%
\caption{Circular velocity functions of satellites and of all galaxies
  in the LG environment. Curves show the cumulative number of subhalos as a function of maximum
  circular velocity, v$_{\rm max}$, averaged over 12 LG volumes at
  resolution level L2. The bottom four curves correspond to subhalos within $300$ kpc of each
  of the two main galaxies; the top two curves to all systems within
  $2$~Mpc from the LG barycenter.  Grey/black curves are from dark
  matter only (DMO) simulations.  Colored curves are for systems that
  contain luminous galaxies in the hydrodynamical runs. The red circles
  show measurements of the most massive MW satellites by Penarrubia et
  al. (2008). For guidance,
  the dashed line denotes a v$_{\rm max}$ of $30$kms$^{-1}$. The
  abundance of satellites with v$_{\rm max}>30$kms$^{-1}$ is halved in
  the hydrodynamic simulations, and matches the MW observations.  At lower values of v$_{\rm max}$, the drop in the
  abundance relative to the DMO case increases as fewer subhalos host an observable galaxy. The inset shows the $2$~Mpc
  curves but for three different resolution runs of the same volume.
  \label{FigVmaxFunc}}
\end{figure}

We next consider the ``too big to fail'' problem
\cite{Boylan-Kolchin-2011}: the observation that the brightest
satellites of the Milky Way appear to shun the most massive dark halo
substructures. A simple statement of the problem is given by the
number of satellite halos with peak circular velocities ($V_{\rm
  max}$) above $\sim 30$ km/s. Only three MW satellites are consistent
with halos more massive than this limit (the two Magellanic Clouds and
the Sagittarius dwarf), whereas {\it dark matter only} (DMO)
$\Lambda$CDM simulations of MW-sized halos \cite{Springel-2008}
produce two to three times this number. Indeed, as shown in
Fig.~\ref{FigVmaxFunc}, when we consider the DMO counterparts of our
LG simulations, the MW and M31 halos each contain an average of $7-8$
satellites with $V_{\rm max}>30$ km/s inside 300 kpc, more than twice
the observed number. This is despite the fact that, in order to match
the most recent dynamical constraints \cite{Gonzalez-2013,
  Penarrubia-2014}, the average halo masses of M31 and the MW in our
simulations are lower than those in which the problem was first
identified \cite{Boylan-Kolchin-2011,Wang-2012}.

The situation changes, however, when we consider the {\it
  hydrodynamical} LG simulations. Compared to the DMO case, the halos
of dwarf galaxies are less massive, with the loss of baryons and a
reduced growth rate leading to a $\sim 15\%$ reduction in $V_{\rm
  max}$. For the abundance of halos below 25~km/s, the reduction in
$V_{\rm max}$ is compounded by the fact that not all low-mass halos
host galaxies: at $10$ km/s the fraction of luminous systems is well
below $10\%$ and decreases even further towards lower masses. Each
main galaxy in our simulation has on average only $3-4$ luminous halos
with $V_{\rm max}>30$~km/s, in excellent agreement with MW
observations \cite{Penarrubia-2008}.

It has also been suggested that constant density cores may be
sufficient and indeed necessary to resolve the ``too big to fail''
problem \cite{Brooks-2013}. Observed stellar kinematics of individual
dwarf spheroidal galaxies have been interpreted as supporting profiles
with profiles\cite{Walker-2011, Amorisco-2012} while others have
argued that they are equally compatible with cusps \cite{Jardel-2013,
  Strigari-2014}. Whereas N-body simulations clearly predict cusps in
CDM halos, it has been shown that supernova feedback may in fact
produce cores \cite{Navarro-1996, Governato-2010, Pontzen-2014},
although perhaps not in most dwarf spheroidals \cite{Penarrubia-2012},
via repeated bursts of star formation in the halo center, resulting
from an assumed high star formation threshold. By contrast, our star
formation and feedback model uses a lower star formation threshold and
produces a realistic galaxy population without cores. Hence, whether
or not cores exist, we conclude that they are not necessary to solve
the perceived small scale problems of $\Lambda$CDM. \\

\begin{figure}
  \resizebox{16cm}{!}{\includegraphics{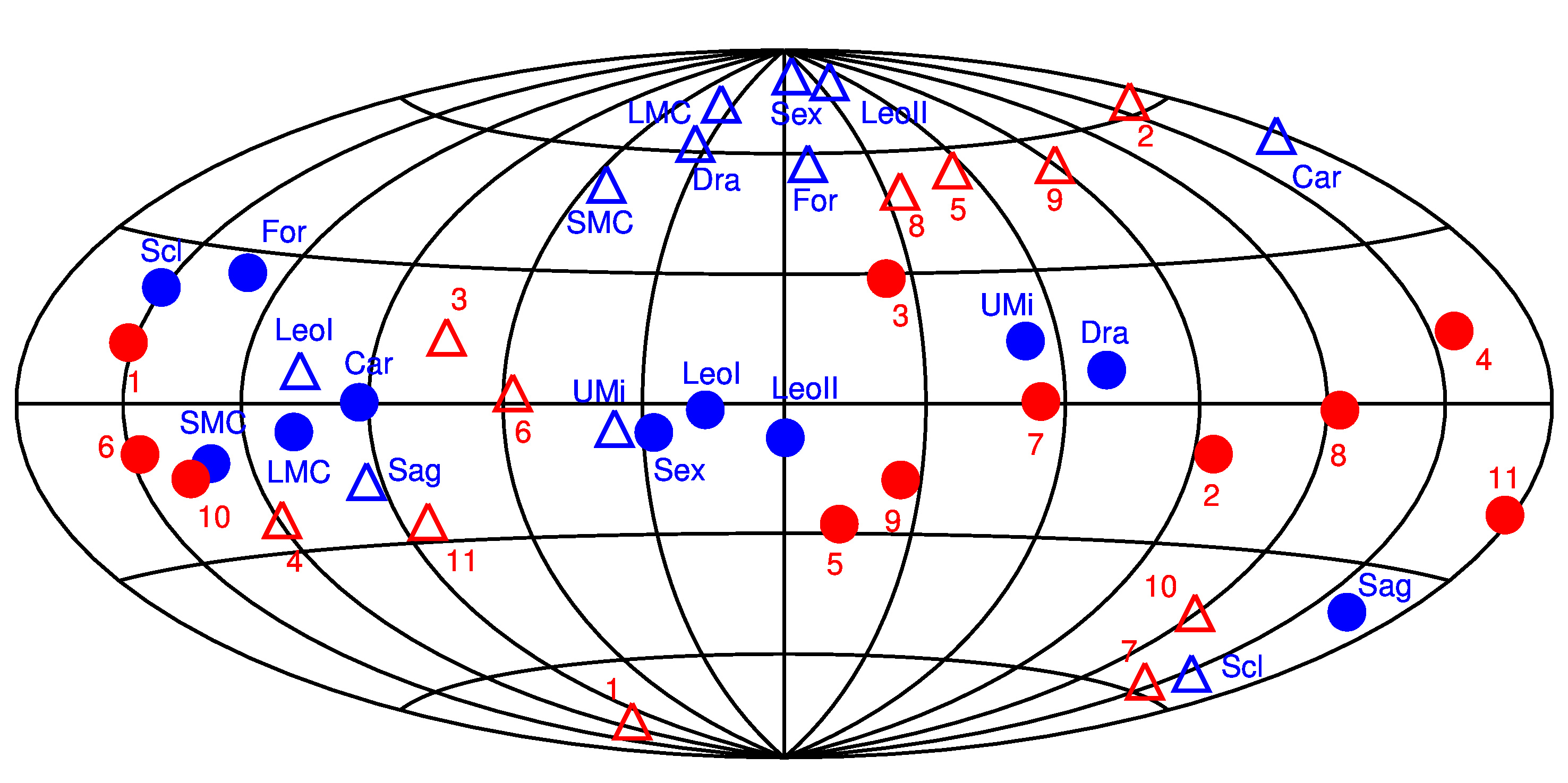}}
\caption{Selecting the brightest satellites, systems as
  anisotropic as the MW's can be formed in $\Lambda$CDM. Red circles show the 
angular distribution of the eleven brightest satellites of a Milky-Way like system
in one of our simulations, while blue circles denote the eleven brightest 
satellites of the Milky Way. Triangles of the
same colors 
indicate the orientation of the corresponding angular momentum
vector. The eleven brightest satellites in the simulated system are
distributed on a plane just as flat as those of the Milky Way and
several of them have a coherent rotation.
\label{FigAnis}}
\end{figure}

Finally, the anisotropy and apparent orbital alignment of the 11
brightest MW satellites, first noticed by
Lynden-Bell \cite{Lynden-Bell-1976}, has been regarded as highly
improbable in $\Lambda$CDM  \cite{Pawlowski-2013}. Fig.~\ref{FigAnis}
compares the observed angular distribution and kinematics to the 11
brightest satellites around one of our resimulated LG galaxies at
$z=0$, as identified in Fig.~\ref{FigLGImage}. Both the simulated and
observed satellite populations show highly anisotropic distributions.

To characterize the anisotropy, we compute the ratios between the
minimal and maximal eigenvalues, $c$ and $a$, of the reduced inertia
tensor \cite{Bailin-2005} defined by the 11 brightest satellites,
${I_{\alpha,\beta} = \sum_{i=1}^{11}{r_{i,\alpha}r_{i,\beta}} /
  r_i^2}$.  We find $\sqrt{c/a}$ in the range 0.34 -- 0.67 in our
simulations, compared to 0.36 for the MW and 0.53 for M31. Clearly,
$\Lambda$CDM can produce satellite systems with a range of
anisotropies, consistent with measurements for both the MW and
M31. The origin of the anisotropy may be traced to the effects of the
``cosmic web'' which impart a degree of coherence to the timing and
direction of satellite accretion \cite{Libeskind-2005}, a scenario
consistent with the recently discovered alignment of satellite planes
with large scale structures \cite{Ibata-2014}.

We conclude that three ``problems'' often cited as inconsistent with
$\Lambda$CDM are resolved in simulations that reproduce the dynamical
constraints of the Local Group environment coupled to a realistic
galaxy formation model. Reionization and feedback allow galaxy
formation to proceed only in a tiny subset of dark matter halos,
eliminating the ``missing satellites'' problem. The loss of baryons
affects the growth of low-mass halos, leading to a reduction in their
maximum circular velocity that solves the ``too big to fail'' problem.
Finally, the structured nature of the $\Lambda$CDM ``cosmic web'' and
galaxy formation within it can lead to highly anisotropic satellite
distributions with correlated kinematics similar to the ``plane of
satellites'' around the Milky Way.

Our results extend the remarkable success of the $\Lambda$CDM model
into a new regime. Future observations may yet throw up new challenges
but to confront them, the theorist's armory will need to include a
realistic account of galaxy formation.

\paragraph*{Acknowledgements}

We are indebted to Dr. Lydia Heck who looks after the supercomputers
at the ICC. This work was supported by the Science and Technology
Facilities Council [grant number ST/F001166/1 and RF040218], the
European Research Council under the European Union’s Seventh Framework
Programme (FP7/2007-2013) / ERC Grant agreement 278594
'GasAroundGalaxies', the National Science Foundation under Grant
No. PHYS-1066293, the Interuniversity Attraction Poles Programme of
the Belgian Science Policy Office [AP P7/08 CHARM]. T. S. acknowledges
the Marie-Curie ITN CosmoComp. C.~S.~F. acknowledges ERC Advanced
Grant 267291 'COSMIWAY' and S.~W. acknowledges ERC Advanced Grant
246797 'GALFORMOD'. This work used the DiRAC Data Centric system at
Durham University, operated by the Institute for Computational
Cosmology on behalf of the STFC DiRAC HPC Facility (www.dirac.ac.uk),
and resources provided by WestGrid (www.westgrid.ca) and Compute
Canada / Calcul Canada (www.computecanada.ca). The DiRAC system is
funded by BIS National E-infrastructure capital grant ST/K00042X/1,
STFC capital grant ST/H008519/1, STFC DiRAC Operations grant
ST/K003267/1, and Durham University. DiRAC is part of the National
E-Infrastructure.

\bibliographystyle{Science}
\bibliography{paper}

\begin{thebibliography}{10}

\bibitem{Klypin-1999}
A.~{Klypin}, A.~V. {Kravtsov}, O.~{Valenzuela}, F.~{Prada}, {\it \apj\/} {\bf
  522}, 82 (1999).

\bibitem{Moore-1999}
B.~{Moore}, {\it et~al.\/}, {\it \apjl\/} {\bf 524}, L19 (1999).

\bibitem{Boylan-Kolchin-2011}
M.~{Boylan-Kolchin}, J.~S. {Bullock}, M.~{Kaplinghat}, {\it \mnras\/} {\bf
  415}, L40 (2011).

\bibitem{Ferrero-2012}
I.~{Ferrero}, M.~G. {Abadi}, J.~F. {Navarro}, L.~V. {Sales}, S.~{Gurovich},
  {\it \mnras\/} {\bf 425}, 2817 (2012).

\bibitem{Pawlowski-2013}
M.~S. {Pawlowski}, P.~{Kroupa}, {\it \mnras\/} {\bf 435}, 2116 (2013).

\bibitem{Lovell-2012}
M.~R. {Lovell}, {\it et~al.\/}, {\it \mnras\/} {\bf 420}, 2318 (2012).

\bibitem{Rocha-2013}
M.~{Rocha}, {\it et~al.\/}, {\it \mnras\/} {\bf 430}, 81 (2013).

\bibitem{Garrison-Kimmel-2014}
S.~{Garrison-Kimmel}, M.~{Boylan-Kolchin}, J.~S. {Bullock}, E.~N. {Kirby}, {\it
  \mnras\/} {\bf 444}, 222 (2014).

\bibitem{Benitez-Llambay-2014}
A.~{Ben{\'{\i}}tez-Llambay}, {\it et~al.\/}, {\it ArXiv e-prints\/}  (2014).

\bibitem{Nuza-2014}
S.~E. {Nuza}, {\it et~al.\/}, {\it \mnras\/} {\bf 441}, 2593 (2014).

\bibitem{Vogelsberger-2014}
M.~{Vogelsberger}, {\it et~al.\/}, {\it \nat\/} {\bf 509}, 177 (2014).

\bibitem{Schaye-2014}
J.~e.~a. {Schaye}, {\it ArXiv e-prints\/}  (2014).

\bibitem{Gonzalez-2013}
R.~E. {Gonzalez}, A.~V. {Kravtsov}, N.~Y. {Gnedin}, {\it ArXiv e-prints\/}
  (2013).

\bibitem{Penarrubia-2014}
J.~{Pe{\~n}arrubia}, Y.-Z. {Ma}, M.~G. {Walker}, A.~{McConnachie}, {\it
  \mnras\/} {\bf 443}, 2204 (2014).

\bibitem{McConnachie-2012}
A.~W. {McConnachie}, {\it \aj\/} {\bf 144}, 4 (2012).

\bibitem{Sawala-2014a}
T.~{Sawala}, {\it et~al.\/}, {\it ArXiv e-prints\/}  (2014).

\bibitem{Springel-2008}
V.~{Springel}, {\it et~al.\/}, {\it \mnras\/} {\bf 391}, 1685 (2008).

\bibitem{Wang-2012}
J.~{Wang}, C.~S. {Frenk}, J.~F. {Navarro}, L.~{Gao}, T.~{Sawala}, {\it
  \mnras\/} {\bf 424}, 2715 (2012).

\bibitem{Penarrubia-2008}
J.~{Pe{\~n}arrubia}, A.~W. {McConnachie}, J.~F. {Navarro}, {\it \apj\/} {\bf
  672}, 904 (2008).

\bibitem{Brooks-2013}
A.~M. {Brooks}, M.~{Kuhlen}, A.~{Zolotov}, D.~{Hooper}, {\it \apj\/} {\bf 765},
  22 (2013).

\bibitem{Walker-2011}
M.~G. {Walker}, J.~{Pe{\~n}arrubia}, {\it \apj\/} {\bf 742}, 20 (2011).

\bibitem{Amorisco-2012}
N.~C. {Amorisco}, N.~W. {Evans}, {\it \mnras\/} {\bf 419}, 184 (2012).

\bibitem{Jardel-2013}
J.~R. {Jardel}, K.~{Gebhardt}, {\it \apjl\/} {\bf 775}, L30 (2013).

\bibitem{Strigari-2014}
L.~E. {Strigari}, C.~S. {Frenk}, S.~D.~M. {White}, {\it ArXiv e-prints\/}
  (2014).

\bibitem{Navarro-1996}
J.~F. {Navarro}, V.~R. {Eke}, C.~S. {Frenk}, {\it \mnras\/} {\bf 283}, L72
  (1996).

\bibitem{Governato-2010}
F.~{Governato}, {\it et~al.\/}, {\it \nat\/} {\bf 463}, 203 (2010).

\bibitem{Pontzen-2014}
A.~{Pontzen}, F.~{Governato}, {\it \nat\/} {\bf 506}, 171 (2014).

\bibitem{Penarrubia-2012}
J.~{Pe{\~n}arrubia}, A.~{Pontzen}, M.~G. {Walker}, S.~E. {Koposov}, {\it
  \apjl\/} {\bf 759}, L42 (2012).

\bibitem{Lynden-Bell-1976}
D.~{Lynden-Bell}, {\it \mnras\/} {\bf 174}, 695 (1976).

\bibitem{Bailin-2005}
J.~{Bailin}, M.~{Steinmetz}, {\it \apj\/} {\bf 627}, 647 (2005).

\bibitem{Libeskind-2005}
N.~I. {Libeskind}, {\it et~al.\/}, {\it \mnras\/} {\bf 363}, 146 (2005).

\bibitem{Ibata-2014}
N.~G. {Ibata}, R.~A. {Ibata}, B.~{Famaey}, G.~F. {Lewis}, {\it \nat\/} {\bf
  511}, 563 (2014).

\bibitem{Springel-2005}
V.~{Springel}, {\it \mnras\/} {\bf 364}, 1105 (2005).

\bibitem{Hopkins-2013}
P.~F. {Hopkins}, {\it \mnras\/} {\bf 428}, 2840 (2013).

\bibitem{Schaye-2010}
J.~{Schaye}, {\it et~al.\/}, {\it \mnras\/} {\bf 402}, 1536 (2010).

\bibitem{Crain-2009}
R.~A. {Crain}, {\it et~al.\/}, {\it \mnras\/} {\bf 399}, 1773 (2009).

\bibitem{Rosas-Guevara-2013}
Y.~M. {Rosas-Guevara}, {\it et~al.\/}, {\it ArXiv 1312.0598\/}  (2013).

\bibitem{Wiersma-2009}
R.~P.~C. {Wiersma}, J.~{Schaye}, B.~D. {Smith}, {\it \mnras\/} {\bf 393}, 99
  (2009).

\bibitem{Haardt-2001}
F.~{Haardt}, P.~{Madau}, {\it Clusters of Galaxies and the High Redshift
  Universe Observed in X-rays\/}, {D.~M.~Neumann \& J.~T.~V.~Tran}, ed. (2001).

\bibitem{Planck-2013}
{Planck Collaboration}, {\it et~al.\/}, {\it ArXiv 1303.5076\/}  (2013).

\bibitem{Wiersma-2009b}
R.~P.~C. {Wiersma}, J.~{Schaye}, T.~{Theuns}, C.~{Dalla Vecchia},
  L.~{Tornatore}, {\it \mnras\/} {\bf 399}, 574 (2009).

\bibitem{Schaye-2000}
J.~{Schaye}, T.~{Theuns}, M.~{Rauch}, G.~{Efstathiou}, W.~L.~W. {Sargent}, {\it
  \mnras\/} {\bf 318}, 817 (2000).

\bibitem{Schaye-2008}
J.~{Schaye}, C.~{Dalla Vecchia}, {\it \mnras\/} {\bf 383}, 1210 (2008).

\bibitem{Schaye-2004}
J.~{Schaye}, {\it \apj\/} {\bf 609}, 667 (2004).

\bibitem{DallaVecchia-2012}
C.~{Dalla Vecchia}, J.~{Schaye}, {\it \mnras\/} {\bf 426}, 140 (2012).

\bibitem{Komatsu-2011}
E.~e.~a. {Komatsu}, {\it \apjs\/} {\bf 192}, 18 (2011).

\bibitem{Flynn-2006}
C.~{Flynn}, J.~{Holmberg}, L.~{Portinari}, B.~{Fuchs}, H.~{Jahrei{\ss}}, {\it
  \mnras\/} {\bf 372}, 1149 (2006).

\bibitem{Tamm-2012}
A.~{Tamm}, E.~{Tempel}, P.~{Tenjes}, O.~{Tihhonova}, T.~{Tuvikene}, {\it
  \aap\/} {\bf 546}, A4 (2012).

\bibitem{Behroozi-2013}
P.~S. {Behroozi}, R.~H. {Wechsler}, C.~{Conroy}, {\it \apj\/} {\bf 770}, 57
  (2013).

\bibitem{Jenkins-2010}
A.~{Jenkins}, {\it \mnras\/} {\bf 403}, 1859 (2010).

\bibitem{Jenkins-2013}
A.~{Jenkins}, {\it \mnras\/} {\bf 434}, 2094 (2013).

\end{thebibliography}

\clearpage
\newpage

\section*{Supplementary Information}

\subsection*{Simulation Code}
The simulations presented in this paper were performed using the code
developed for the {\it Evolution and Assembly of GaLaxies and their
  Environments} project ({\sc Eagle} \cite{Schaye-2014}). The {\sc
  Eagle} code is a substantially modified version of {\sc P-Gadget-3},
which itself is an improved version of the publicly available {\sc
  Gadget-2} code \cite{Springel-2005}. Gravitational accelerations are
computed using a Tree-PM scheme, while hydrodynamic forces are
computed using smoothed particle hydrodynamics (SPH) in the
pressure-entropy formalism introduced by Hopkins \cite{Hopkins-2013}.

{\sc Eagle} is an evolution of the models used in the {\sc
  Owls} \cite{Schaye-2010} and {\sc Gimic} \cite{Crain-2009} projects
and has been calibrated to reproduce the $z=0$ stellar mass function
accurately from $10^8\Ms$ to $10^{11}\Ms$ in a cosmological volume of
$100^3$ Mpc$^3$. In addition, the {\sc Eagle} code also successfully
reproduces many other properties and scaling laws of observed galaxy
populations, including the evolution of the stellar mass function, and
the luminosities, colors and morphologies of galaxies on many
different scales.

The subgrid physics model of {\sc Eagle} is described in detail by
Schaye {\it et al.} \cite{Schaye-2014}. It includes radiative cooling,
star formation, stellar evolution and stellar mass loss, and thermal
feedback that captures the collective effects of stellar winds,
radiation pressure and supernova explosions. It also includes black
hole growth fueled by gas accretion and mergers, and feedback from AGN
\cite{Rosas-Guevara-2013}.

Following Wiersma {\it et al.} \cite{Wiersma-2009}, net cooling rates
are computed separately for 11 elements, assuming ionization
equilibrium in the presence of uniform UV and X-ray backgrounds from
quasars and galaxies \cite{Haardt-2001}, and the cosmic microwave
background (CMB) \cite{Planck-2013}. Hydrogen is assumed to reionize
instantaneously at $z=11.5$, which is implemented by turning on the
ionizing background. At higher redshifts the background is truncated
at 1 Ryd, limiting its effect to preventing the formation of molecular
hydrogen. During reionization an extra 2 eV per proton mass are
injected to account for the increase in the photoheating rates of
optically thick gas over the optically thin rates that are used
otherwise. For hydrogen this is done at $z=11.5$, ensuring that the
gas is quickly heated to $10^4$~K, but for HeII the extra heat is
distributed in time following a Gaussian centered at $z=3.5$ with
$\sigma(z) = 0.5$, which reproduces the observed thermal history
\cite{Wiersma-2009b, Schaye-2000}.

The star formation rate is assumed to be
pressure-dependent \cite{Schaye-2008} and follows the observed
Kennicutt-Schmidt star formation law with a metallicity-dependent
density threshold \cite{Schaye-2004}. Energy feedback from star
formation is implemented using the stochastic, thermal prescription of
Dalla Vecchia \& Schaye \cite{DallaVecchia-2012}. The expectation value
for the energy injected per unit stellar mass formed decreases with
the metallicity of the gas and increases with the gas density to
account for unresolved radiative losses and to help prevent spurious,
numerical losses. The injected energy is calibrated to reproduce the
observed, $z=0$ galaxy stellar mass function. On average it is close
to the energy available from core collapse supernovae
alone \cite{Schaye-2014}. Galactic winds develop naturally, without
imposing mass loading factors, velocities or directions.

The three different resolution levels of our LG simulations labelled
``L1'', ``L2'' and ``L3'' have primordial gas (DM) particle masses of
$1.0 (5.0) \times 10^4 \Ms$, $1.3 (5.9) \times10^5 \Ms $ and $1.5
(7.3) \times10^6 \Ms$, respectively, and maximum gravitational
softening lengths of 94 pc, 216 pc and 500 pc. L3 is close to the
resolution of the {\sc Eagle} L100N1504 simulation. While the
resolution in our highest resolution simulations is comparable to the
best simulations of individual MW sized galaxies, it should be noted
that with a gravitational softening length of 94 pc, they barely
resolve the scale height of the MW thin disk. We cannot truly resolve
individual star forming regions, and rely instead on a well-calibrated
subgrid physics model to parametrize the star formation and feedback
processes. While there is clearly scope for future improvements, the
stellar mass function and the circular velocity function of
substructures in our simulations are well converged at L2. This
indicates that our numerical resolution is sufficient to capture the
physical mechanisms of structure formation, gas accretion and
outflows, and that the assumptions made in the {\sc Eagle} subgrid
model do not alter the results significantly.

The high resolution regions always enclose a sphere of at least
2.5~Mpc radius from the LG barycenter at $z=0$.  Outside the high
resolution regions, dark matter particles of increasing mass are used
to sample the large scale environment of the $100^3$~Mpc$^3$ parent
simulation. To investigate the impact of baryons, we also repeated all
our simulations as dark matter only (DMO), where the dark matter
particle masses in the high resolution region are larger by a factor
of $(\Omega_b+\Omega_{DM})/\Omega_{DM}$ than in the corresponding
hydrodynamic simulations.

For our study, we have used the same parameter values that were used
in the $100^3$Mpc$^3$ L100N1504 {\sc Eagle}
simulation \cite{Schaye-2014} independently of resolution. We have
found very good convergence in all the relevant properties of our
simulated galaxies. While the {\it Eagle} simulations use the Planck-1
cosmology \cite{Planck-2013}, the Local Group simulations were
performed using the WMAP-7 cosmology \cite{Komatsu-2011}, with density
parameters at $z=0$ for matter, baryons and dark energy of
$\Omega_{M}=0.272$, $\Omega_b=0.0455$ and $\Omega_\lambda=0.728$,
respectively, a Hubble parameter of H$_0=70.4$kms$^{-1}$Mpc$^{-1}$, a
power spectrum of (linear) amplitude on the scale of 8$h^{-1}$Mpc of
$\sigma_8=0.81$ and a power-law spectral index $n_s=0.967$.

\subsection*{The Local Group Simulations}
Our twelve Local Group regions are zoom simulations based on a DMO
simulation (called {\sc ``DOVE''}) of $100^3$Mpc$^3$ with $1620^3$
particles in the WMAP-7 cosmology. The resimulation volumes were
selected to match the dynamical constraints of the Local Group. Each
volume contains a pair of halos in the mass range $5\times10^{11} \Ms$
to $2.5\times10^{12} \Ms$, with median values of $1.4
\times10^{12}\Ms$ for the primary (more massive) halo and $0.9
\times10^{12}\Ms$ for the secondary (less massive) halo of each
pair. The combined masses of the primary and secondary range from
$1.6\times10^{12}\Ms$ to $3.6\times10^{12}\Ms$ with a median mass of
$2.3\times10^{12}\Ms$, in good agreement with recent estimates of
$2.40_{-1.05}^{+1.95}\times10^{12}\Ms$ based on dynamical arguments
and CDM simulations \cite{Gonzalez-2013}, or $2.3 \pm 0.7 \times
10^{12}$ based on equations of motions that take into account the
observed velocities of galaxies in the local
volume \cite{Penarrubia-2014}. We further require that the two halos be
separated by $800 \pm 200$ kpc, approaching with radial velocity of
$\left(0-200\right)$ kms$^{-1}$ and with tangential velocity below
$100$ kms$^{-1}$; to have no additional halo larger than the smaller
of the pair within 2.5 Mpc from the LG barycenter, and to be in
environments with a relatively unperturbed Hubble flow out to 4 Mpc.

Excluding substructures, the stellar masses of the Milky Way and M31
analogues in our simulations lie in the range $1.5 - 5.5 \times
10^{10} \Ms$, comparable to the observational estimates for the Milky
Way ($5 \times 10^{10} \Ms$ \cite{Flynn-2006}) but lower than those for
M31 ($10^{11} \Ms$ \cite{Tamm-2012}). It should be noted, however, that
the Milky Way and M31 both appear to lie above the average
stellar-to-halo mass relation derived from abundance
matching \cite{Behroozi-2013}, which is well reproduced by our model in
a large cosmological volume \cite{Schaye-2014}. While the predicted
abundance of satellites and dwarf galaxies within the Local Group
varies with the total mass density, which is dominated by dark matter,
it is largely independent of the stellar mass of the main
galaxies. The main galaxies in almost all our simulations are
extended disk galaxies, reminiscent of the Milky Way and M31, as can
be seen for one example in Fig.~1.

The high resolution initial conditions were created using second-order
Lagrangian perturbation theory \cite{Jenkins-2010}. The cosmological
parameters and the linear phases of {\sc Dove}, which are taken from
the public multi-scale Gaussian white noise field {\sc Panphasia}, are
given in Tables 1 and 6 by Jenkins \cite{Jenkins-2013}, which also
describes the method used to make the Local Group zoom initial
conditions.

\clearpage

\end{document}